\documentclass[conference]{IEEEtran}
\ifCLASSINFOpdf
\else
\fi
\newcommand*{\affaddr}[1]{#1} % No op here. Customize it for different styles.
\newcommand*{\affmark}[1][*]{\textsuperscript{#1}}
\newcommand*{\email}[1]{\texttt{#1}}
\usepackage{graphicx}

\begin{document}
\title{KPynq: A Work-Efficient Triangle-Inequality \\ based K-means on FPGA}

\author{

Yuke Wang\affmark[1], Zhaorui Zeng\affmark[2], Boyuan Feng\affmark[1], Lei Deng\affmark[2], and Yufei Ding\affmark[1]\\
\affaddr{\affmark[1]Department of Computer Science}\\
\affaddr{\affmark[2]Department of Electrical and Computer Engineering}\\
\email{\affmark[1]\{yuke\_wang,boyuan,yufeiding\}@cs.ucsb.edu} \\ 
\email{\affmark[2]\{zzeng00,leideng\}@ucsb.edu}\\
\affaddr{University of California, Santa Barbara}

}


%Telephone: (800) 555--1212, Fax: (888) 555--1212}


\maketitle

\begin{abstract}
K-means is a popular but computation-intensive algorithm for unsupervised learning. To address this issue, we present KPynq, a work-efficient triangle-inequality based K-means on FPGA for handling large-size, high-dimension datasets. KPynq leverages an algorithm-level optimization to balance the performance and computation irregularity, and a hardware architecture design to fully exploit the pipeline and parallel processing capability of various FPGAs. In the experiment, KPynq consistently outperforms the CPU-based standard K-means in terms of its speedup (up to 4.2$\times$) and significant energy-efficiency (up to 218$\times$).
\end{abstract}


% \fi
\IEEEpeerreviewmaketitle

\section{Introduction}
\begin{figure} \small
    \centering
    \includegraphics[width=0.49\textwidth, clip=true]{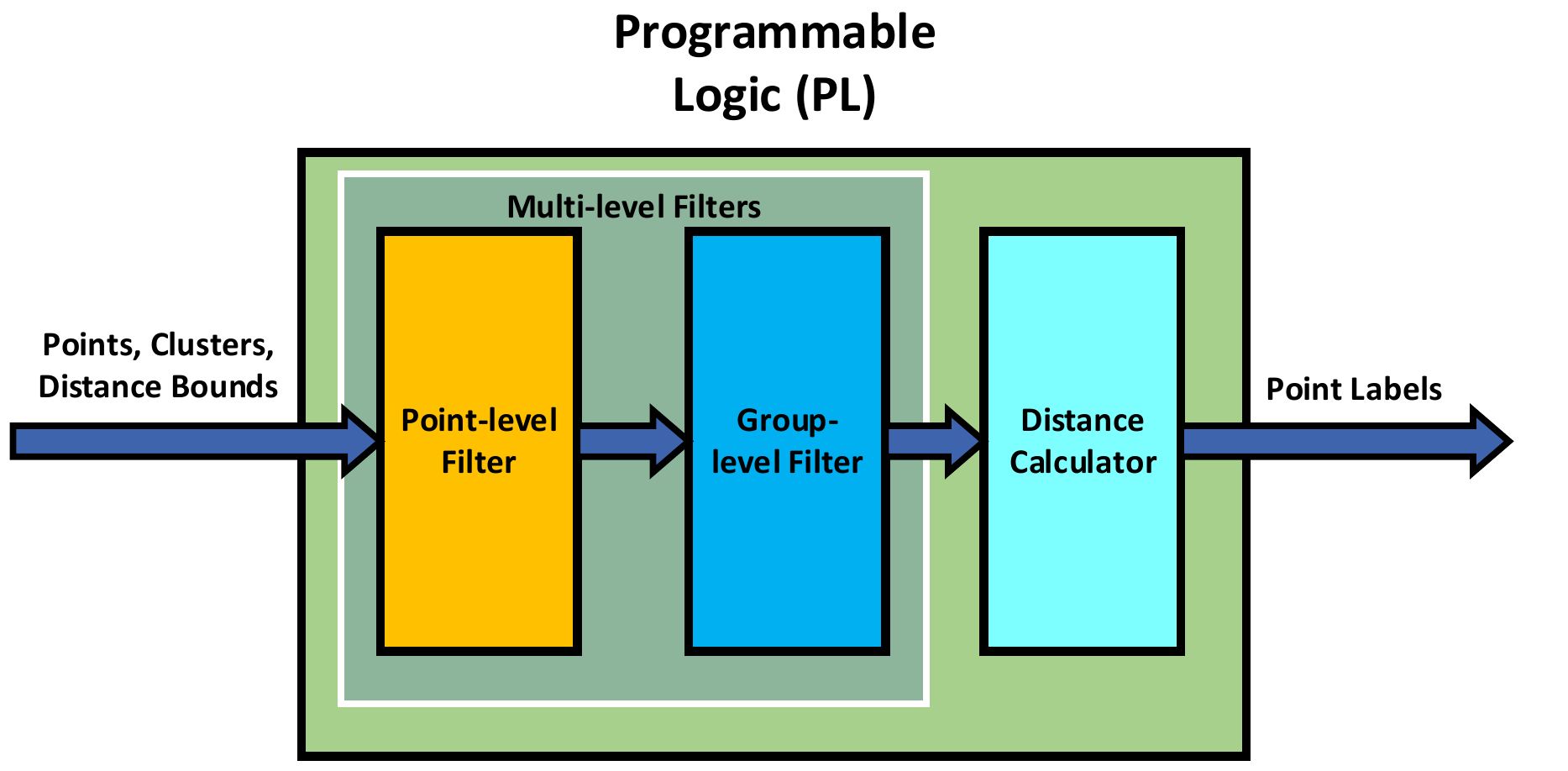}
    \caption{Overview of KPynq.}
    \label{fig: Overall Architecture}
\end{figure}

K-means   clustering is a widely  applied  unsupervised  learning algorithms,  finding  its strength in many machine learning application scenarios, such as  unlabeled  data  clustering,  image  segmentation,  and  feature  learning. Despite  its  popularity,  standard  K-means  usually has  unsatisfactory  performance  due  to  its  high  computation complexity. Previous research studies in K-means hardware acceleration~\cite{1232813} \cite{5963944} optimize K-means for the specific dataset or certain FPGA, which lack adaptability and flexibility. However, KPynq is much more scalable and highly configurable equipped with a set of tunable parameters (e.g. degree of parallelism), which help to handle various datasets. KPynq is targeted at Pynq-Z1, which is based on Xilinx Zynq SoC \cite{websitepynq}. This SoC consists of two subsystems: PS (Processing System) and PL (Programmable Logic). Besides, a DMA controller and a high-performance AXIS streaming interface build the data connection between PS and PL. A Python program in PS is responsible for invoking the PL part hardware accelerator and initiate the DMA data transfer. The PL part hardware accelerator of KPynq, as shown in Fig. \ref{fig: Overall Architecture}, includes two main components: Multi-level Filters (Point-level and Group-level Filter) and Distance Calculator. Multi-level Filters is for reducing distance computations at the algorithmic level, while the Distance Calculator is for doing distance computations which have not been filtered out.

\section{Experiment and Conclusion}
Our KPynq design is implemented by using the Xilinx Vivado Design Suite v2018.2. and is deployed on Pynq-Z1 board \cite{websitepynq}. This board is built on ZYNQ XC7Z020-1CLG400C all-programmable SoC, which has a 650 MHz dual-core ARM Cortex-A9 processor (PS) and an Artix-7 family programmable logic (PL) on the same die. Each Cortex-A9 processor core has 32 KB L1 4-way cache and shares a 512 KB L2 cache with other cores. The programmable logic has 13,300 logic slices, each with four 6-input LUTs and 8 flip-flops, 630 KB BRAM (280 BRAM\_18K), and 220 DSP slices. The auxiliary parts used by our design include a DMA controller and AXIS buses for the data communication among PS, PL, and external DRAM. Experiments show that KPynq consistently excels an optimized CPU-based standard K-means implementation with $2.95\times$ speedup, and $150.90\times$ better energy-efficiency on average across the six real-life datasets from \cite{Dua:2017}, which covers a wide range of size and dimensionality. 

\bibliographystyle{IEEEtran}
\bibliography{bibliography.bib}

\begin{thebibliography}{1}
\providecommand{\url}[1]{#1}
\csname url@samestyle\endcsname
\providecommand{\newblock}{\relax}
\providecommand{\bibinfo}[2]{#2}
\providecommand{\BIBentrySTDinterwordspacing}{\spaceskip=0pt\relax}
\providecommand{\BIBentryALTinterwordstretchfactor}{4}
\providecommand{\BIBentryALTinterwordspacing}{\spaceskip=\fontdimen2\font plus
\BIBentryALTinterwordstretchfactor\fontdimen3\font minus
  \fontdimen4\font\relax}
\providecommand{\BIBforeignlanguage}[2]{{%
\expandafter\ifx\csname l@#1\endcsname\relax
\typeout{** WARNING: IEEEtran.bst: No hyphenation pattern has been}%
\typeout{** loaded for the language `#1'. Using the pattern for}%
\typeout{** the default language instead.}%
\else
\language=\csname l@#1\endcsname
\fi
#2}}
\providecommand{\BIBdecl}{\relax}
\BIBdecl

\bibitem{1232813}
A.~G.~S. Filho, A.~C. Frery, C.~C. de~Araujo, H.~Alice, J.~Cerqueira, J.~A.
  Loureiro, M.~E. de~Lima, M.~G.~S. Oliveira, and M.~M. Horta, ``Hyperspectral
  images clustering on reconfigurable hardware using the k-means algorithm,''
  in \emph{16th Symposium on Integrated Circuits and Systems Design, 2003.
  SBCCI 2003. Proceedings.}, Sep. 2003, pp. 99--104.

\bibitem{5963944}
H.~M. Hussain, K.~Benkrid, H.~Seker, and A.~T. Erdogan, ``Fpga implementation
  of k-means algorithm for bioinformatics application: An accelerated approach
  to clustering microarray data,'' in \emph{2011 NASA/ESA Conference on
  Adaptive Hardware and Systems (AHS)}, June 2011, pp. 248--255.

\bibitem{websitepynq}
\BIBentryALTinterwordspacing
``Pynq-z1 reference manual [reference.digilentinc].'' [Online]. Available:
  \url{https://reference.digilentinc.com/reference/programmable-logic/pynq-z1/reference-manual}
\BIBentrySTDinterwordspacing

\bibitem{Dua:2017}
\BIBentryALTinterwordspacing
D.~Dheeru and E.~K. Taniskidou, ``{UCI} machine learning repository,'' 2017.
  [Online]. Available: \url{http://archive.ics.uci.edu/ml}
\BIBentrySTDinterwordspacing

\end{thebibliography}

\end{document}